\documentclass[prb,aps,showpacs,floatfix,floats,nobalancelastpage,twocolumn]{revtex4}
\usepackage{amsmath, euscript}
\usepackage{amsfonts}
\usepackage{amssymb}
\usepackage{graphicx}
\usepackage{color}
\usepackage{dcolumn}
\usepackage{bm}
\usepackage{subfigure}
\usepackage{ulem}
\usepackage{CJKutf8}

\def\etal{{\it et~al}}
\def\ra{\rangle}
\def\la{\langle}
\def\Hc{{\rm H.c.}}

\begin{document}

\title{Variational study of $J_1$--$J_2$ Heisenberg model on Kagome lattice using projected Schwinger boson wave functions}

\author{Tiamhock Tay (\begin{CJK}{UTF8}{gbsn}郑添福\end{CJK})}
\author{Olexei I. Motrunich}

\affiliation{Department of Physics, California Institute of Technology, Pasadena, CA 91125}

\date{\today}



\begin{abstract}
  Motivated by the unabating interest in the spin-1/2 Heisenberg antiferromagnetic model on the Kagome lattice, we investigate the energetics of projected Schwinger boson (SB) wave functions in the $J_1$--$J_2$ model with antiferromagnetic $J_2$ coupling.  Our variational Monte Carlo results show that Sachdev's $Q_1=Q_2$ SB ansatz has a lower energy than the Dirac spin liquid for $J_2\gtrsim 0.08 J_1$ and the ${\bf q=0}$ Jastrow type magnetically ordered state.  This work demonstrates that the projected SB wave functions can be tested on the same footing as their fermionic counterparts.
\end{abstract}
\maketitle


The Heisenberg antiferromagnet on the kagome lattice has long been anticipated to realize a spin liquid (SL). Recently, Herbertsmithite ${\rm ZnCu_3(OH)_6Cl_2}$,\cite{Helton2006,Ofer2006,Mendels2007,Olariu2007} which contains kagome layers of spin-1/2 moments, has emerged as an experimental candidate with no sign of any ordering down to 50 mK.   Interest in spin liquids on the kagome lattice has been re-ignited by recent works,\cite{Jiang2008,Sindzingre2009,Yan2010,Lauchli2011} where Density Matrix Renormalization Group (DMRG) studies find a spin disordered ground state with a small gap.  A vast review of earlier literature is revisited in Ref.~\onlinecite{Yan2010}.  Preliminary DMRG data suggests that the system moves deeper into the spin liquid phase upon adding small antiferromagnetic second-neighbor $J_2$ coupling,\cite{White2010} and this supports earlier Exact Diagonalization (ED) study that found an increase in the gap for $J_2$ up to $0.1$.\cite{Sindzingre2009}

In an early study of the nearest-neighbor model using large-$N$ treatment of Schwinger boson (SB) slave particles, Sachdev\cite{Sachdev1992} found that condensation of spinons gives rise to magnetically ordered ground states for spin $S>0.26$.  A subsequent variational study by Sindzingre~\etal\cite{Sindzingre1994} using Resonating Valence Bond (RVB) wave functions interpolating between spin liquids and magnetically ordered states instead found that the former have lower energies. However, the spin correlations beyond first-neighbors do not agree well with ED results.\cite{Sindzingre1994,Leung1993} More recently, the trial energy of the projected Dirac spin liquid constructed using fermionic slave particle approach was found to lie very close to the ED ground state energy in the nearest-neighbor Heisenberg model,\cite{Hastings2000,Ying2007} and a very recent study extended this to the presence of second-neighbor coupling $J_2$.\cite{Iqbal2010} However, the observation of an energy gap in the DMRG and ED studies suggests that the gapless Dirac spin liquid may lose some ground in the presence of $J_2$. \cite{Sindzingre2009,White2010}

\begin{figure}
  \centering
  \begin{tabular}{cc}
    \subfigure[~${\bf q=0}$ SB ansatz]{
      \includegraphics[trim=0 -10 0 0, scale=0.4]{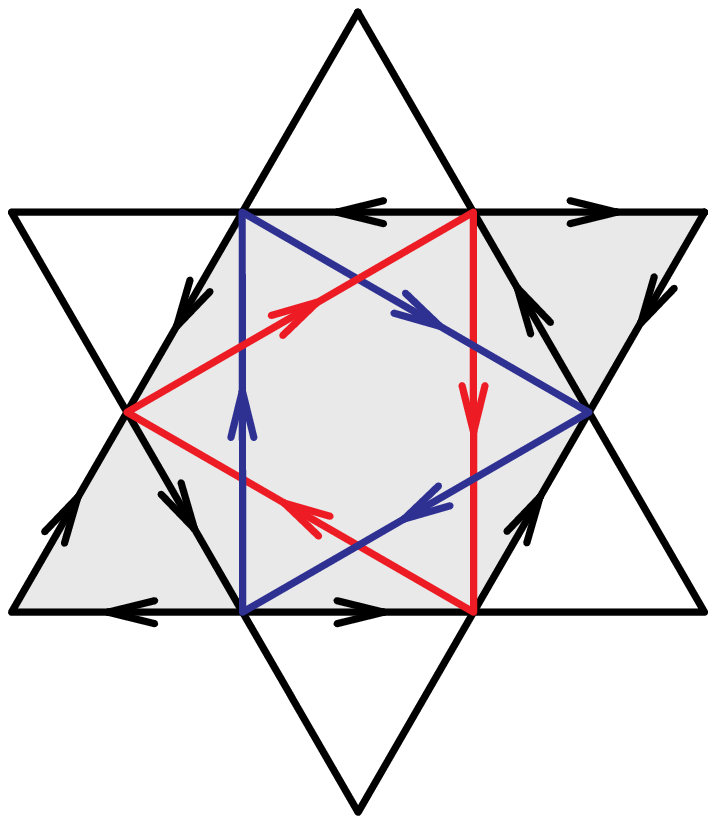}
      \label{fig:ansatz2}
    } &
    \subfigure[~$\sqrt{3}\times\sqrt{3}$ SB ansatz]{
      \includegraphics[trim=0 -10 0 0, scale=0.4]{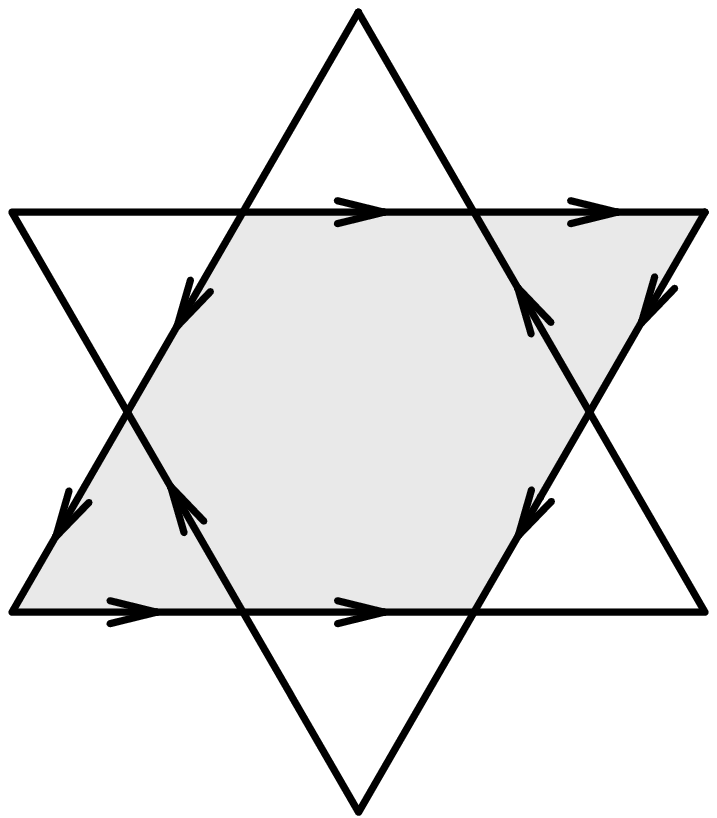}
      \label{fig:ansatz1}
    }
  \end{tabular}
  \vskip -2mm
  \caption{The SB ansatze $\{A_{ij}\}$ from Ref.~\onlinecite{Sachdev1992}. The unit cell is shaded for each ansatz. All equidistant $A_{ij}$ have identical magnitudes, and $A_{ij}$ is positive if an arrow points from site $i$ to $j$. For the ${\bf q=0}$ ansatz, we extend $A_{ij}$ to include second-neighbor pairing.  The $\sqrt{3}~\times\sqrt{3}$ ansatz has poorer energy for $J_2 > 0$.
}
  \label{fig:ansatz}
\end{figure}

To investigate this possibility, we study the energetics of a class of projected Schwinger boson wave functions in the $J_1$--$J_2$ Heisenberg model with Hamiltonian
\begin{eqnarray}
  \hat{H} = J_1\sum_{\la i,j\ra} {\bf S}_i \cdot {\bf S}_j + J_2\sum_{\la\la i,j\ra\ra} {\bf S}_i \cdot {\bf S}_j,
\end{eqnarray}
where $\la i,j\ra$ and $\la\la i,j\ra\ra$ denotes first and second-neighbor pairs.   We set $J_1=1$ as the unit for energy and consider only antiferromagnetic $J_2\geq 0$.    The Schwinger-boson representation of a spin ${\bf S}$ is given by\cite{Auerbach}
\begin{equation}
  {\bf S} = \frac{1}{2}\sum_{\sigma,\sigma'}b^\dagger_\sigma {\bf\mbox{\boldmath$\sigma$}}_{\sigma\sigma'} b_{\sigma'}, ~~~\kappa = \sum_\sigma b^\dagger_\sigma b_\sigma = 2S,
\end{equation}
where $b_\sigma$ is a bosonic operator, ${\bf\mbox{\boldmath$\sigma$}}$ are Pauli matrices and $\kappa$ is the number of bosons per site. The Hamiltonian becomes quartic in the bosonic operators, and upon mean-field decoupling\cite{Auerbach, Sachdev1992, Wang2006} leads to the following
\begin{eqnarray}
  \hat{H}_{\rm m.f.} &=& \frac{1}{2}\sum_{i,j} \left(A_{ij} b^\dagger_{i\downarrow} b^\dagger_{j\uparrow} + \Hc\right) + \sum_{i,j}\frac{\vert A_{ij}\vert^2}{2J_{ij}}\nonumber \\
  &-&\mu \sum_i \left(\sum_\sigma b^\dagger_{i\sigma} b_{i\sigma}-\kappa\right),\\
  A_{ij} &=& \frac{1}{2} J_{ij} \sum_{\sigma,\sigma'} \epsilon_{\sigma\sigma'}\la b_{i\sigma}b_{j\sigma'}\ra, ~~~\kappa = \sum_\sigma \la b^\dagger_{i\sigma} b_{i\sigma}\ra.~\label{eq:self_consistency}
\end{eqnarray}
We treat the ``pairing amplitudes'' $A_{ij}$ and ``chemical potential'' $\mu$ as variational parameters. Equations~(\ref{eq:self_consistency}) are self-consistency relations in the mean field.    Figure~\ref{fig:ansatz} shows Sachdev's ansatze for $A_{ij}$ which have good mean field energies in the $J_1$-only model.\cite{Sachdev1992}     The nearest-neighbor $A_{ij}$ are real, and their signs are positive if an arrow points from site $i$ to $j$.   Following Sindzingre~\etal,\cite{Sindzingre1994} we label these as ${\bf q=0}$ and $\sqrt{3}\times\sqrt{3}$ Schwinger boson ansatze, which correspond to Sachdev's $Q_1=Q_2$ and $Q_1=-Q_2$ respectively\cite{Sachdev1992, Huh2010} (also called $[\pi{\rm Hex},0{\rm Rhom}]$ and $[0{\rm Hex},0{\rm Rhom}]$ in Refs.~\onlinecite{Wang2006,Messio2010}).  For the ${\bf q=0}$ SB ansatz shown in Fig.~\ref{fig:ansatz2}, we introduce an additional real parameter for the second-neighbor pairing, with the pattern of arrows going clockwise in triangular loops for both first and second neighbors.   Second-neighbor $A_{ij}$ for the $\sqrt{3}\times\sqrt{3}$ SB ansatz are forbidden by its projective symmetry group (PSG).   In Ref.~\onlinecite{Wang2006}, Wang~\etal~ found two more distinct ansatze for symmetric spin liquids, but they argued that the Heisenberg model energies of those spin liquids are expected to be considerably poorer. Our variational calculations confirm that this is indeed true.

\begin{figure}
  \centering
  \includegraphics[width=\columnwidth]{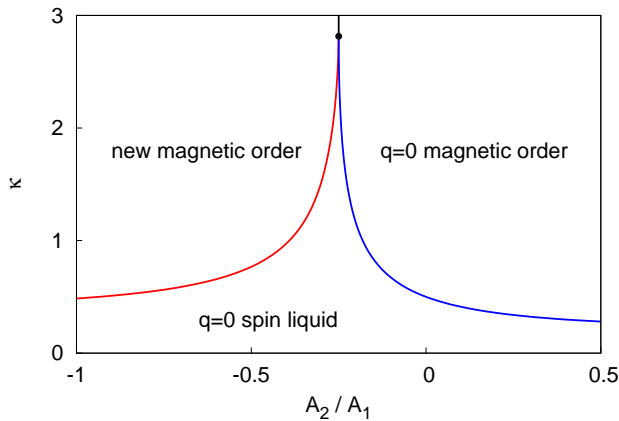}
  \vskip -2mm
  \caption{Mean field ``phase diagram'' for the ${\bf q=0}$ SB ansatz.\cite{MFphaseDiagram} The phase boundary separates spin liquid from the magnetically ordered phases.  A spin liquid regime is present for physical spin-1/2 systems with $\kappa$=1. The new magnetic order is complex and is not relevant for this paper.}
  \label{fig:density}
\end{figure}

We first present the results of a crude study on the accessibility of Schwinger boson spin liquids at the mean field level, by computing the critical boson density
\begin{equation}
  \kappa_c = -1 + \frac{1}{N}\sum_{\vert\lambda_\alpha\vert\neq\vert\mu_{\rm max}\vert} \frac{\vert\mu_{\rm max}\vert}{\sqrt{\mu_{\rm max}^2-\lambda_\alpha^2}},\label{eq:density}
\end{equation}
accessible in the mean field without Bose condensation in the thermodynamic limit. 
Here $\{\lambda_\alpha\}$ are eigenvalues of the matrix $-i\hat{A}$ [cf.\ Eq.~(\ref{eq:diagonalization}) below], $\mu_{\rm max}=-{\max}\{\vert\lambda_\alpha\vert\}$, and $N$ is the number of sites on the lattice.    Below $\kappa_c$, the mean field excitation spectrum is gapped and gives rise to a stable spin liquid. For $\kappa\geq\kappa_c$, the gap closes and magnetic ordering results from spinon condensation.  With only nearest-neighbor $A_{ij}$, $\kappa_c\approx 0.5$ and $0.54$ for the ${\bf q=0}$ and $\sqrt{3}\times\sqrt{3}$ SB ansatze respectively.\cite{Sachdev1992, Wang2006}  In analogy to Wang's analysis for the honeycomb lattice,\cite{Wang2010} Fig.~\ref{fig:density} shows the critical boson density $\kappa_c$ versus $A_2/A_1$ for the ${\bf q=0}$ SB ansatz, where $A_1$ and $A_2$ are the amplitudes of first and second-neighbor $A_{ij}$.  We note that $\kappa_c > 1$ in the parameter range $-0.4<A_2/A_1<-0.18$, i.e., the second-neighbor pairing has opened up a disordered regime relevant for $S=1/2$.\cite{MFphaseDiagram}

\begin{figure}
  \centering
  \begin{tabular}{cc}
    \subfigure[~${\bf q=0}$ MO]{
      \includegraphics[trim=0 -10 0 0, scale=0.37]{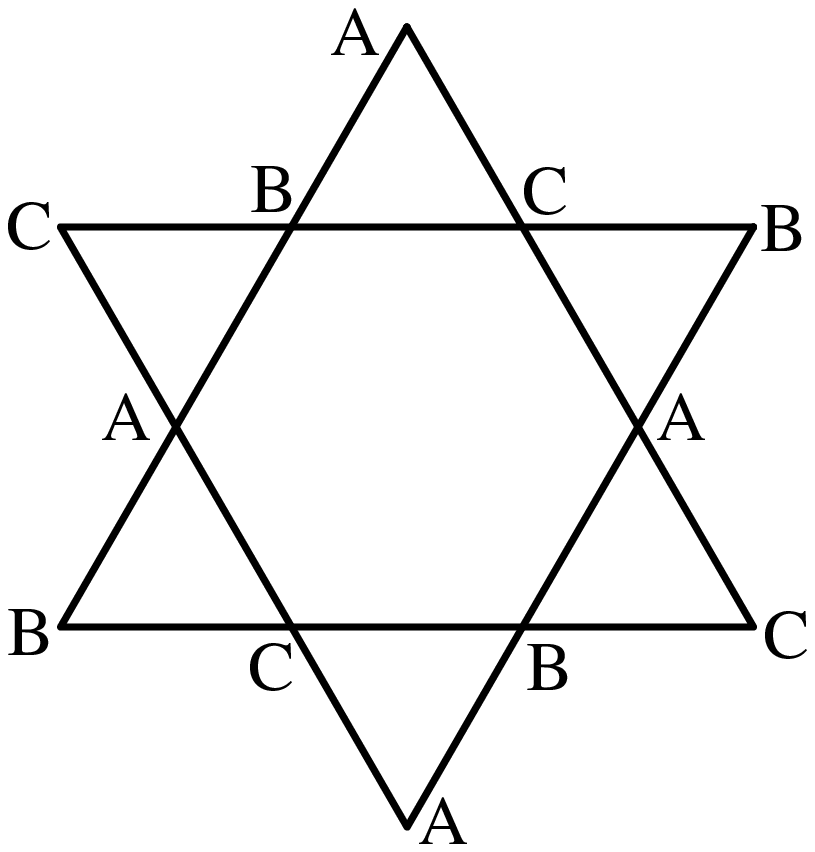}
      \label{fig:order2}
    } &
    \subfigure[~$\sqrt{3}\times\sqrt{3}$ MO]{
      \includegraphics[trim=0 -10 0 0, scale=0.37]{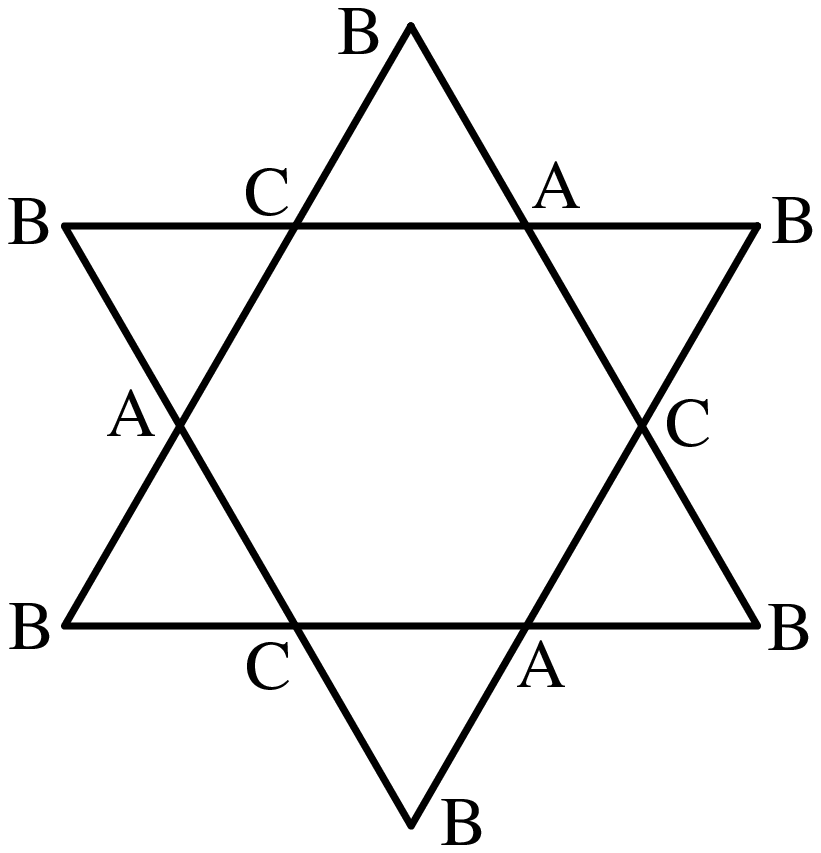}
      \label{fig:order1}
    }
  \end{tabular}
  \vskip -2mm
  \caption{Magnetic orderings (MO) which arise from spinon condensation in ${\bf q=0}$ and $\sqrt{3}\times\sqrt{3}$ SB ansatze. A, B, and C are the 120$^\circ$ antiferromagnetic spin orientations. For $J_2>0$, the $\sqrt{3}~\times\sqrt{3}$ MO has poorer energy than the ${\bf q=0}$ MO.}
  \label{fig:magnetic_order}
\end{figure}

We now turn to a variational Monte Carlo (VMC) study of the $J_1$--$J_2$ model on the symmetric 36-site cluster used in previous numerical studies,\cite{Leung1993,Sindzingre1994} which allows a direct comparison with ED energies as well as the energies of the Dirac SL and magnetically ordered states. We construct projected SB wave functions as follows. Writing the real anti-symmetric matrix $\hat{A}$ [see Eq.~(\ref{eq:self_consistency})] as
\begin{equation}
  \hat{A} = i\hat{M}\hat{\Lambda} \hat{M}^\dagger,\label{eq:diagonalization}
\end{equation}
where $\hat{\Lambda}$ is diagonal and $\hat{M}$ is unitary, we solve the mean field Hamiltonian using Bogoliubov's transformation and obtain the following trial wave function\cite{Auerbach}
\begin{eqnarray}
  \vert\Psi_{\rm SB}\ra &=& \hat{\cal P}_{G} ~\exp \left\{\sum_{j,k}u_{jk} ~b^\dagger_{j\uparrow} b^\dagger_{k\downarrow} \right\} \vert 0\ra,\label{eq:projectedWF}\\
  u_{jk} &=& i\sum_\alpha \frac{M_{j\alpha} \lambda_\alpha (M^\dagger)_{\alpha k}}{-\mu + \sqrt{\mu^2-\lambda_\alpha^2}}.\label{eq:matrix_elements}
\end{eqnarray}
The Gutzwiller operator $\hat{\cal P}_G$ enforces the constraint $\kappa=1$ at every site.   Although this density may not be accessible to a given ansatz and $\mu$ at the mean field level (see Fig.~\ref{fig:density}), the projected SB wave function is a valid variational state in the physical Hilbert space.  Here, $u_{jk}$ decays exponentially for $\mu<\mu_{\rm max}$ and the projected SB wave function realizes a short-range RVB state when written in the valence bond basis.  In the limit $\mu\ll \mu_{\rm max}$, Eq.~(\ref{eq:matrix_elements}) shows that the pattern of $u_{jk}$ roughly follows that of $A_{jk}$. More generally, the PSG of the ${\bf q=0}$ SB ansatz enforces the pattern of $u_{jk}$ postulated in Sindzingre~\etal,\cite{Sindzingre1994} where $u_{jk}$ only connect sites on the different ``sublattices" A,B,C defined in the sense of the ${\bf q=0}$ magnetic order in Fig.~\ref{fig:magnetic_order}(a).  At $\mu = \mu_{\rm max}$, $u_{jk}$ decays in a power-law $\sim \vert r_j-r_k\vert^{-3}$; one can view $\mu\rightarrow\mu_{\rm max}$ as a finite size realization of magnetic orders.\cite{Auerbach, Messio2010} 

In our VMC simulations, the amplitude of each sampled spin configuration is given by the permanent of the $N/2\times N/2$ matrix $\{u_{jk}\}$, where $j$ and $k$ run over the spin-up and spin-down sites respectively.  We make use of Ryser-Nijenhuis-Wilf dense permanent algorithm to calculate the permanent;\cite{permanent} in this way, the procedure does not have the sign problem encountered in the valence bond basis.\cite{Sindzingre1994}  Despite a poor computational cost scaling $\sim 2^{N/2}$ with system size, it is manageable for sizes that are already interesting.    

We also consider magnetically ordered states shown in Fig.~\ref{fig:magnetic_order} which arise from the condensation of spinons in the respective ansatz.\cite{Sachdev1992} For both orderings, their classical nearest-neighbor energies are identical, but the second-neighbor energy is clearly lower for the ${\bf q=0}$ ordered state since it has antiferromagnetic second-neighbor correlations while the $\sqrt{3}\times\sqrt{3}$ state has ferromagnetic correlations. It is therefore sufficient to consider only the former.   We construct the following trial wave function
\begin{equation}
  \la\{S^z_j\}\vert\Psi^{\rm MO}_{\bf q=0} \ra = \exp\left\{i \sum_j\phi_j S^z_j -\sum_{ij}K_{ij}S^z_i S^z_j\right\},
\end{equation}
where $\phi_j=\{0,\pm2\pi/3\}$ are the phase angles on the three sublattices in Fig.~\ref{fig:magnetic_order}(a), and $K_{ij}$ are two-body pseudo-potentials for the Jastrow factor.  Such Jastrow-type wave functions are widely used in VMC studies due to their simplicity.  We allow two variational parameters for the first and second-neighbor pseudo-potentials, and two more for a power-law decay between further neighbors.  We also consider a Huse-Elser\cite{Huse1988} type of three-site phase factor allowed by the symmetry of the classical state, but it apparently does not improve the trial energy.

\begin{figure}
  \centering
  \includegraphics[width=\columnwidth]{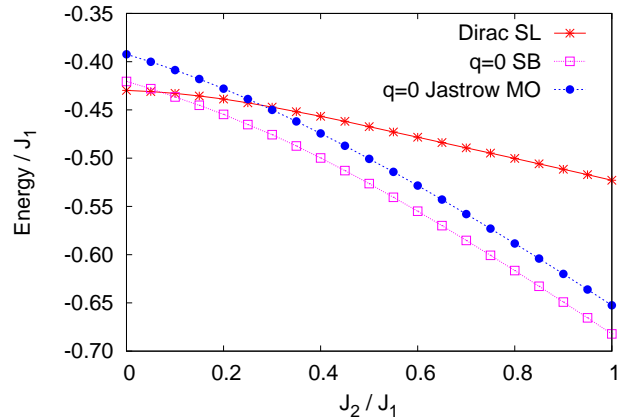}
  \vskip -2mm
  \caption{Comparison of trial energies per site for Dirac SL, ${\bf q=0}$ SB wave function, and ${\bf q=0}$ Jastrow-type magnetically ordered (MO) state. The SB state has poorer energy than Dirac SL for $J_2/J_1\lesssim 0.08$, but performs better for larger $J_2$ and better than the Jastrow-type MO for all $J_2$.}
  \label{fig:energies}
\end{figure}

To get an idea of the variational energetics landscape, we also include the Dirac SL constructed from fermionic spinons hopping with flux $\pi$ through hexagons and flux $0$ through elementary triangles.\cite{Hastings2000,Ying2007}  This state was extended in Ref.~\onlinecite{Iqbal2010} to include second-neighbor hopping such that triangles formed by two nearest-neighbor bonds and one second-neighbor bond have flux $\pi$.   The amplitude of the second-neighbor hopping provides a single variational parameter.  For large sizes, we reproduce results in Ref.~\onlinecite{Iqbal2010} for $J_2>0$; we perform VMC for the present 36-site cluster and find negligible size dependence on the scale in Fig.~\ref{fig:energies}.

Figure~\ref{fig:energies} shows the variational energies of the Dirac SL, the ${\bf q=0}$ SB wave function, and the ${\bf q=0}$ magnetically ordered state. For the $J_1$-only model, the Dirac SL has significantly better energy than the SB state.  However, the latter improves quickly with $J_2 > 0$, and becomes lowest for $J_2\gtrsim 0.08$ among the wave functions in this study. The ${\bf q=0}$ Jastrow-type MO state has higher energy for all $J_2$ values shown. 

\begin{figure}
  \centering
  \includegraphics[width=\columnwidth]{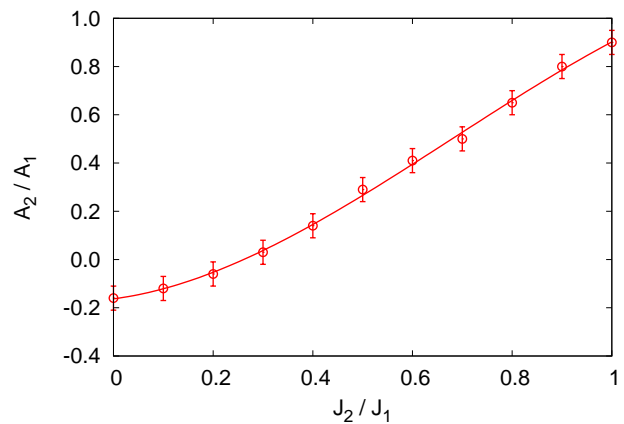}
  \vskip -2mm
  \caption{Optimal $A_2/A_1$ versus $J_2/J_1$ for the ${\bf q=0}$ projected SB wave function, also optimized over $\mu$ for each $J_2$.}
  \label{fig:A2-J2}
\end{figure}

The ${\bf q=0}$ SB wave function has two variational parameters, $A_2$ and $\mu$.  For $J_2=0$, the lowest energy of $-0.420$ per site occurs at $A_2\approx -0.15$.   Interestingly, this $A_2$ approaches the spin liquid window in Fig.~\ref{fig:density} where we observe fairly narrow spinon bands.   We find antiferromagnetic correlations between second-neighbor sites. These results are very close to those obtained by Sindzingre~\etal ~in a variational study of the $J_1$-only model,\cite{Sindzingre1994} wherein they considered RVB ansatze with a few variational parameters for nearby-neighbors $u_{jk}$ and a power-law decay $\sim \vert r_j-r_k\vert^{-p}$ with $p=5$ for further neighbors. For our projected state, $\mu$ optimizes very close to $\mu_{\rm max}$ which corresponds to formal $p=3$ in the thermodynamic limit.  Despite the difference in the details of the realizations, both are suggestive of a near-critical state at $J_2=0$.  Thus, Table~V in Ref.~\onlinecite{Sindzingre1994} indicates that such wave functions have significant ${\bf q=0}$ correlations across the full 36-site cluster.

Figure~\ref{fig:A2-J2} shows the optimal $A_2$ against the second-neighbor coupling $J_2$.  We find that $A_2$ increases with $J_2$ and is important for improving the trial energy of the ${\bf q=0}$ SB wave function.  Beyond $J_2\sim 0.1$, the optimal $\mu$ starts to decrease away from $\mu_{\rm max}$, e.g., it is $1.02 \mu_{\rm max}$ for $J_2=0.2$ and moves further to $1.05 \mu_{\rm max}$ for $J_2=0.4$.

{\it Discussion.} Our energetics study reveals the ${\bf q=0}$ SB wave function as a viable candidate for the $J_1$--$J_2$ Heisenberg model.  This is perhaps not surprising since this state is quite competitive in the $J_1$-only model\cite{Sindzingre1994} and has antiferromagnetic second-neighbor correlations which are favorable when $J_2>0$ is added.  The ${\bf q=0}$ SB state can furthermore accommodate the $J_2$ coupling by varying $A_2$. In the large $J_2$ limit, the system breaks into three independent Kagome networks, each as difficult as the original nearest-neighbor Kagome problem.  The large-$A_2$ state in the large $J_2$ limit is just like the $A_1$-only state in the $J_1$-only model, so while not the best, is again reasonably good in energy.   Thus, the $A_1$--$A_2$ ansatz provides a way to interpolate between the small $J_2$ and large $J_2$ regimes and is an appealing candidate. It wins against the Dirac SL and against the best Jastrow wave function for the ${\bf q=0}$ MO, but it may also correspond to possible MO at intermediate $J_2$. It would be very interesting to check our results against exact calculations in the $J_1$--$J_2$ model on the 36-site cluster to assess the accuracy of the projected ${\bf q=0}$ SB state.

From our Schwinger boson wave function study, we cannot address the question whether the ground state is spin liquid or has magnetic long range order.  It is known\cite{Auerbach,Sindzingre1994} that the RVB wave functions can realize both phases depending on the range of the valence bond amplitudes $u_{jk}$.  In the projected SB wave function setup, if $\mu$ is very close to $\mu_{\rm max}$, this can be viewed as a finite-size realization of the spinon condensation and hence magnetic order.  On the other hand, if $\mu$ is a finite distance away from $\mu_{\rm max}$, this gives exponentially decaying $u_{jk}$ and hence short-range RVB spin liquid. While we find that $\mu$ optimizes away from $\mu_{\rm max}$ for $J_2 > 0.1$, these small-size results cannot be used to establish the long-distance behavior, and the ultimate phase determination must come from exact studies on larger systems.  Nevertheless, we hope that our demonstration of the viability of the ${\bf q=0}$ SB wave function\cite{Sindzingre1994, Messio2010, Huh2010, Lauchli2009} can be useful for further studies of the $J_1$--$J_2$ Kagome antiferromagnet.

In this work we used a dense permanent routine;\cite{permanent} such calculations can in principle be pursued to 48 sites.  If we also restrict $u_{jk}$ to only few nearby-neighbors, the VMC can be scaled further due to sparseness of the matrix.  Simulations in the valence bond basis may reach larger sizes;\cite{Sindzingre1994}  attention to the sign problem is needed there although it is less severe than the sign problems in QMC.  An important aspect of our work is the demonstration that projected Schwinger boson wave functions can be tested on the same footing as the slave fermion spin liquids, for smaller but still reasonable system sizes, and can be included in the VMC toolbox. Here we highlight our use of permanents in a variational study of Heisenberg model on the triangular lattice in magnetic field,\cite{Tay2010} where we obtained excellent wave functions for Mott insulators and supersolids of bosons with frustrated hopping.  We suggest the honeycomb spin liquid\cite{Meng2010} and Wang's proposal\cite{Wang2010} as one context for applying the projected SB wave functions, as well as other model proposals in Ref.~\onlinecite{Wang2006} for realizing new spin liquids on the triangular and kagome lattices.\cite{Wang2006}

{\it Acknowledgments}: OIM would like to acknowledge the KITP program ``Disentangling Quantum Many-body Systems: Computational and Conceptual Approaches'' and thank many participants for discussions, and in particular to Steven White for the inspiring informal rerun of his talk and for sharing new results. The research is supported by the NSF through grant DMR-0907145 and the A.~P.~Sloan Foundation.

\bibliography{biblio}

\begin{thebibliography}{26}
\expandafter\ifx\csname natexlab\endcsname\relax\def\natexlab#1{#1}\fi
\expandafter\ifx\csname bibnamefont\endcsname\relax
  \def\bibnamefont#1{#1}\fi
\expandafter\ifx\csname bibfnamefont\endcsname\relax
  \def\bibfnamefont#1{#1}\fi
\expandafter\ifx\csname citenamefont\endcsname\relax
  \def\citenamefont#1{#1}\fi
\expandafter\ifx\csname url\endcsname\relax
  \def\url#1{\texttt{#1}}\fi
\expandafter\ifx\csname urlprefix\endcsname\relax\def\urlprefix{URL }\fi
\providecommand{\bibinfo}[2]{#2}
\providecommand{\eprint}[2][]{\url{#2}}

\bibitem[{\citenamefont{Helton et~al.}(2007)}]{Helton2006}
\bibinfo{author}{\bibfnamefont{J.~S.} \bibnamefont{Helton}}
  \bibnamefont{et~al.}, \bibinfo{journal}{Phys. Rev. Lett.}
  \textbf{\bibinfo{volume}{98}}, \bibinfo{pages}{107204}
  (\bibinfo{year}{2007}).

\bibitem[{\citenamefont{Ofer et~al.}(2006)}]{Ofer2006}
\bibinfo{author}{\bibfnamefont{O.}~\bibnamefont{Ofer}} \bibnamefont{et~al.},
  \bibinfo{journal}{cond-mat/0610540}  (\bibinfo{year}{2006}).

\bibitem[{\citenamefont{Mendels et~al.}(2007)}]{Mendels2007}
\bibinfo{author}{\bibfnamefont{P.}~\bibnamefont{Mendels}} \bibnamefont{et~al.},
  \bibinfo{journal}{Phys. Rev. Lett.} \textbf{\bibinfo{volume}{98}},
  \bibinfo{pages}{077204} (\bibinfo{year}{2007}).

\bibitem[{\citenamefont{Olariu et~al.}(2008)}]{Olariu2007}
\bibinfo{author}{\bibfnamefont{A.}~\bibnamefont{Olariu}} \bibnamefont{et~al.},
  \bibinfo{journal}{Phys. Rev. Lett.} \textbf{\bibinfo{volume}{100}},
  \bibinfo{pages}{087202} (\bibinfo{year}{2008}).

\bibitem[{\citenamefont{Jiang et~al.}(2008)\citenamefont{Jiang, Weng, and
  Sheng}}]{Jiang2008}
\bibinfo{author}{\bibfnamefont{H.~C.} \bibnamefont{Jiang}},
  \bibinfo{author}{\bibfnamefont{Z.~Y.} \bibnamefont{Weng}}, \bibnamefont{and}
  \bibinfo{author}{\bibfnamefont{D.~N.} \bibnamefont{Sheng}},
  \bibinfo{journal}{Phys. Rev. Lett.} \textbf{\bibinfo{volume}{101}},
  \bibinfo{pages}{117203} (\bibinfo{year}{2008}).

\bibitem[{\citenamefont{Sindzingre and Lhuillier}(2009)}]{Sindzingre2009}
\bibinfo{author}{\bibfnamefont{P.}~\bibnamefont{Sindzingre}} \bibnamefont{and}
  \bibinfo{author}{\bibfnamefont{C.}~\bibnamefont{Lhuillier}},
  \bibinfo{journal}{EPL (Europhysics Letters)} \textbf{\bibinfo{volume}{88}},
  \bibinfo{pages}{27009} (\bibinfo{year}{2009}).

\bibitem[{\citenamefont{Yan et~al.}(2010)\citenamefont{Yan, Huse, and
  White}}]{Yan2010}
\bibinfo{author}{\bibfnamefont{S.}~\bibnamefont{Yan}},
  \bibinfo{author}{\bibfnamefont{D.~A.} \bibnamefont{Huse}}, \bibnamefont{and}
  \bibinfo{author}{\bibfnamefont{S.~R.} \bibnamefont{White}},
  \bibinfo{journal}{arXiv:1011.6114}  (\bibinfo{year}{2010}).

\bibitem[{\citenamefont{{L{\"a}uchli} et~al.}(2011)\citenamefont{{L{\"a}uchli},
  {Sudan}, and {S{\o}rensen}}}]{Lauchli2011}
\bibinfo{author}{\bibfnamefont{A.~M.} \bibnamefont{{L{\"a}uchli}}},
  \bibinfo{author}{\bibfnamefont{J.}~\bibnamefont{{Sudan}}}, \bibnamefont{and}
  \bibinfo{author}{\bibfnamefont{E.~S.} \bibnamefont{{S{\o}rensen}}},
  \bibinfo{journal}{arXiv:1103.1159}  (\bibinfo{year}{2011}).

\bibitem[{Whi()}]{White2010}
\bibinfo{note}{S.~R.~White and D.~A.~Huse [private communication].}

\bibitem[{\citenamefont{Sachdev}(1992)}]{Sachdev1992}
\bibinfo{author}{\bibfnamefont{S.}~\bibnamefont{Sachdev}},
  \bibinfo{journal}{Phys. Rev. B} \textbf{\bibinfo{volume}{45}},
  \bibinfo{pages}{12377} (\bibinfo{year}{1992}).

\bibitem[{\citenamefont{Sindzingre et~al.}(1994)\citenamefont{Sindzingre,
  Lecheminant, and Lhuillier}}]{Sindzingre1994}
\bibinfo{author}{\bibfnamefont{P.}~\bibnamefont{Sindzingre}},
  \bibinfo{author}{\bibfnamefont{P.}~\bibnamefont{Lecheminant}},
  \bibnamefont{and}
  \bibinfo{author}{\bibfnamefont{C.}~\bibnamefont{Lhuillier}},
  \bibinfo{journal}{Phys. Rev. B} \textbf{\bibinfo{volume}{50}},
  \bibinfo{pages}{3108} (\bibinfo{year}{1994}).

\bibitem[{\citenamefont{Leung and Elser}(1993)}]{Leung1993}
\bibinfo{author}{\bibfnamefont{P.~W.} \bibnamefont{Leung}} \bibnamefont{and}
  \bibinfo{author}{\bibfnamefont{V.}~\bibnamefont{Elser}},
  \bibinfo{journal}{Phys. Rev. B} \textbf{\bibinfo{volume}{47}},
  \bibinfo{pages}{5459} (\bibinfo{year}{1993}).

\bibitem[{\citenamefont{Hastings}(2000)}]{Hastings2000}
\bibinfo{author}{\bibfnamefont{M.~B.} \bibnamefont{Hastings}},
  \bibinfo{journal}{Phys. Rev. B} \textbf{\bibinfo{volume}{63}},
  \bibinfo{pages}{014413} (\bibinfo{year}{2000}).

\bibitem[{\citenamefont{Ran et~al.}(2007)\citenamefont{Ran, Hermele, Lee, and
  Wen}}]{Ying2007}
\bibinfo{author}{\bibfnamefont{Y.}~\bibnamefont{Ran}},
  \bibinfo{author}{\bibfnamefont{M.}~\bibnamefont{Hermele}},
  \bibinfo{author}{\bibfnamefont{P.~A.} \bibnamefont{Lee}}, \bibnamefont{and}
  \bibinfo{author}{\bibfnamefont{X.-G.} \bibnamefont{Wen}},
  \bibinfo{journal}{Phys. Rev. Lett.} \textbf{\bibinfo{volume}{98}},
  \bibinfo{pages}{117205} (\bibinfo{year}{2007}).

\bibitem[{\citenamefont{Iqbal et~al.}(2010)\citenamefont{Iqbal, Becca, and
  Poilblanc}}]{Iqbal2010}
\bibinfo{author}{\bibfnamefont{Y.}~\bibnamefont{Iqbal}},
  \bibinfo{author}{\bibfnamefont{F.}~\bibnamefont{Becca}}, \bibnamefont{and}
  \bibinfo{author}{\bibfnamefont{D.}~\bibnamefont{Poilblanc}},
  \bibinfo{journal}{arXiv:1011.3954}  (\bibinfo{year}{2010}).

\bibitem[{Aue()}]{Auerbach}
\bibinfo{note}{A. Auerbach, {\it Interacting Electrons and Quantum Magnetism},
  Springer-Verlag, Berlin (1994).}

\bibitem[{\citenamefont{Wang and Vishwanath}(2006)}]{Wang2006}
\bibinfo{author}{\bibfnamefont{F.}~\bibnamefont{Wang}} \bibnamefont{and}
  \bibinfo{author}{\bibfnamefont{A.}~\bibnamefont{Vishwanath}},
  \bibinfo{journal}{Phys. Rev. B} \textbf{\bibinfo{volume}{74}},
  \bibinfo{pages}{174423} (\bibinfo{year}{2006}).

\bibitem[{\citenamefont{Huh et~al.}(2010)\citenamefont{Huh, Fritz, and
  Sachdev}}]{Huh2010}
\bibinfo{author}{\bibfnamefont{Y.}~\bibnamefont{Huh}},
  \bibinfo{author}{\bibfnamefont{L.}~\bibnamefont{Fritz}}, \bibnamefont{and}
  \bibinfo{author}{\bibfnamefont{S.}~\bibnamefont{Sachdev}},
  \bibinfo{journal}{Phys. Rev. B} \textbf{\bibinfo{volume}{81}},
  \bibinfo{pages}{144432} (\bibinfo{year}{2010}).

\bibitem[{\citenamefont{Messio et~al.}(2010)\citenamefont{Messio, C\'epas, and
  Lhuillier}}]{Messio2010}
\bibinfo{author}{\bibfnamefont{L.}~\bibnamefont{Messio}},
  \bibinfo{author}{\bibfnamefont{O.}~\bibnamefont{C\'epas}}, \bibnamefont{and}
  \bibinfo{author}{\bibfnamefont{C.}~\bibnamefont{Lhuillier}},
  \bibinfo{journal}{Phys. Rev. B} \textbf{\bibinfo{volume}{81}},
  \bibinfo{pages}{064428} (\bibinfo{year}{2010}).

\bibitem[{MFp()}]{MFphaseDiagram}
\bibinfo{note}{The minimum of the spinon dispersion is at ${\bf k=0}$ for
  $A_2/A_1 > -0.25$ and jumps to new locations for $A_2/A_1 < -0.25$. The
  relevant spinon bands happen to be very narrow near the transition and this
  results in large $\kappa_c$.}

\bibitem[{\citenamefont{Wang}(2010)}]{Wang2010}
\bibinfo{author}{\bibfnamefont{F.}~\bibnamefont{Wang}}, \bibinfo{journal}{Phys.
  Rev. B} \textbf{\bibinfo{volume}{82}}, \bibinfo{pages}{024419}
  (\bibinfo{year}{2010}).

\bibitem[{per()}]{permanent}
\bibinfo{note}{I.~M.~Wanless, {\it Permanents}, Chapter 31 in {\it Handbook of
  Linear Algebra} (ed. L. Hogben), Chapman \& Hall/CRC (2007).}

\bibitem[{\citenamefont{Huse and Elser}(1988)}]{Huse1988}
\bibinfo{author}{\bibfnamefont{D.~A.} \bibnamefont{Huse}} \bibnamefont{and}
  \bibinfo{author}{\bibfnamefont{V.}~\bibnamefont{Elser}},
  \bibinfo{journal}{Phys. Rev. Lett.} \textbf{\bibinfo{volume}{60}},
  \bibinfo{pages}{2531} (\bibinfo{year}{1988}).

\bibitem[{\citenamefont{{L{\"a}uchli} and {Lhuillier}}(2009)}]{Lauchli2009}
\bibinfo{author}{\bibfnamefont{A.}~\bibnamefont{{L{\"a}uchli}}}
  \bibnamefont{and}
  \bibinfo{author}{\bibfnamefont{C.}~\bibnamefont{{Lhuillier}}},
  \bibinfo{journal}{arXiv:0901.1065}  (\bibinfo{year}{2009}).

\bibitem[{\citenamefont{Tay and Motrunich}(2010)}]{Tay2010}
\bibinfo{author}{\bibfnamefont{T.}~\bibnamefont{Tay}} \bibnamefont{and}
  \bibinfo{author}{\bibfnamefont{O.~I.} \bibnamefont{Motrunich}},
  \bibinfo{journal}{Phys. Rev. B} \textbf{\bibinfo{volume}{81}},
  \bibinfo{pages}{165116} (\bibinfo{year}{2010}).

\bibitem[{\citenamefont{Meng et~al.}(2010)\citenamefont{Meng, Lang, Wessel,
  Assaad, and Muramatsu}}]{Meng2010}
\bibinfo{author}{\bibfnamefont{Z.~Y.} \bibnamefont{Meng}},
  \bibinfo{author}{\bibfnamefont{T.~C.} \bibnamefont{Lang}},
  \bibinfo{author}{\bibfnamefont{S.}~\bibnamefont{Wessel}},
  \bibinfo{author}{\bibfnamefont{F.~F.} \bibnamefont{Assaad}},
  \bibnamefont{and}
  \bibinfo{author}{\bibfnamefont{A.}~\bibnamefont{Muramatsu}},
  \bibinfo{journal}{Nature} \textbf{\bibinfo{volume}{464}},
  \bibinfo{pages}{847} (\bibinfo{year}{2010}).

\end{thebibliography}

\end{document}